\documentclass[]{spiejour}  
\usepackage{graphicx} 
\usepackage{amsmath}
\usepackage{amsfonts}
\usepackage{amssymb}
\usepackage{epsfig}
\usepackage{color}
\usepackage{subcaption}
\usepackage{lscape}
\usepackage{epstopdf}
\usepackage {color}
\usepackage{multirow}
\usepackage{setspace}
\usepackage{cite}
\usepackage[usenames,dvipsnames,table]{xcolor}
\usepackage[colorlinks=true, allcolors=blue]{hyperref}
\usepackage{mathtools}
\usepackage{tocloft}

\DeclarePairedDelimiter{\ceil}{\lceil}{\rceil}

\def\cyan{\textcolor{black}}
\definecolor{lightgray}{gray}{0.8}
\definecolor{llightgray}{gray}{0.95}

\def\##1{{\bf #1}}
\def\=#1{\underline{\underline{#1}}}

\def\*#1{\breve{{\bf #1}}}
\def\+#1{\breve{\underline{\underline #1}}}

\def\le{\left(}
\def\ri{\right)}
\def\les{\left[}
\def\ris{\right]}
\def\lec{\left\{}
\def\ric{\right\}}

\def\.{\mbox{ \tiny{$^\bullet$} }}

\def\eps{\varepsilon}

\def\Nbar{\ensuremath{\bar{N}}}
\def\nopt{n_{\rm opt}}

\def\O{\scriptscriptstyle 0}
\def\epso{\eps_{\scriptscriptstyle 0}}
 
\def\lambdao{\lambda_{\scriptscriptstyle 0}}
\def\muo{\mu_{\scriptscriptstyle 0}}

\def\etao{\eta_{\scriptscriptstyle 0}}

\def\co{c_{\scriptscriptstyle 0}}

\def\lambdaomin{\lambda_{\O_{min}}}
\def\lambdaomax{\lambda_{\O_{max}}}

\def\sfE{{\sf E}}

\def\ux{\hat{\underline x}}
\def\uy{\hat{\underline y}}
\def\uz{\hat{\underline z}}

\def\param{\xi}
\def\InGaN{\ensuremath{\mbox{In}_\param\mbox{Ga}_{1\mbox{\tiny-}\param}\mbox{N}}}
\def\InGaNn{\ensuremath{\mbox{In}_\param\mbox{Ga}_{1\mbox{\tiny-}\param}\mbox{N}}}
\def\JscOpt{\ensuremath{J_{SC}^{Opt}}}
\def\mAcm2{mA~cm\ensuremath{^{-2}}}

\title{Optimization of nonhomogeneous indium-gallium-nitride Schottky-barrier thin-film solar cells}

\author{Tom H. Anderson\supscr{1}, Akhlesh Lakhtakia\supscr{2,$\ast$}, and
Peter~B.~Monk\supscr{1} }

\affiliation{
$^1$University of Delaware, Department of Mathematical Sciences, 501 Ewing Hall,  Newark, DE 19716, USA\\
$^2$Pennsylvania State University, Department of Engineering Science and Mechanics, \\ NanoMM---Nanoengineered Metamaterials Group, 212 EES Building, \\ University Park, PA 16802, USA\\
$^\ast$\linkable{akhlesh@psu.edu}
}

\cftpagenumbersoff{figure}
\cftpagenumbersoff{table} 

\begin{document}
\maketitle

\begin{abstract}

A two-dimensional model was developed to simulate the optoelectronic characteristics of indium-gallium-nitride (\InGaN), thin-film, 
\cyan{Schottky-barrier  solar} cells. The solar cells comprise a window, designed to reduce the reflection of incident light, Schottky-barrier and ohmic front {electrodes}, an $n$-doped \InGaN~wafer, and a metallic periodically corrugated  back-reflector (PCBR).  The  ratio of indium to gallium in the wafer varies periodically throughout the thickness of \cyan{the absorbing layer
of} the solar cell. Thus, the resulting \InGaN~wafer's optical and electrical properties are made to vary periodically. This material nonhomogeneity could be physically achieved by varying the fractional composition of indium and gallium during deposition. Empirical models for indium nitride and gallium nitride were combined using Vegard\rq{}s law to determine the optical and electrical constitutive properties of the alloy. The   nonhomogeneity of the electrical properties of the \InGaN~aids in the separation of the excited electron--hole pairs, while  the periodicities of optical properties and the back-reflector enable the incident light to couple to multiple guided wave modes. The profile of the resulting charge-carrier-generation rate  when the solar cell is illuminated by the AM1.5G spectrum was calculated using the rigorous coupled-wave approach. The steady-state drift-diffusion equations were solved using COMSOL, which employs finite-volume methods, to calculate the current density as a function of the voltage. Mid-band Shockley--Read--Hall, Auger, and radiative recombination rates were taken to be the dominant methods of recombination. The model was used to study the effects of the solar-cell geometry and the shape of the periodic  material nonhomogeneity on efficiency. The solar-cell efficiency was optimized using the differential evolution algorithm.
\end{abstract}


\section{Introduction} \label{intro}
Alloys of indium gallium nitride (\InGaNn) can be tailored to possess a wide range of bandgaps, from 0.70~eV to 3.42~eV, by varying the relative proportions of indium and gallium through the parameter $\xi \in (0,1)$ \cite{Wu2003}. Pure indium nitride (i.e., $\xi=1$) has a bandgap of $~0.7$~eV \cite{Wu2002, Saito2002}, whereas gallium nitride (i.e., $\xi=0$) has a bandgap of $3.42$~eV. It should be noted {that \InGaN~with} high indium content (i.e., $\xi \gtrsim 0.3$) currently suffers from poor electrical characteristics, background $n$-doping due to Fermi pinning above the conduction-band edge \cite{Li2005}, and a bandgap that is greater than expected\cite{Osamura1972,Yodo2002}. These problems are exacerbated by $p$-doping {of  \InGaN~\cite{Hamady2016}.}

Solar cells can be designed to use an in-built potential provided by a Schottky-barrier junction, which can occur at a {metal/semiconductor} interface \cite{Swami,Colinge}. By partnering $n$-doped \InGaNn~with 
a metal possessing a large work function \cyan{$\Phi$}---as opposed to, say, employing the more usual $p$-$i$-$n$ junction---the problems associated with $p$-doping of the material are avoided. Furthermore, the deposition process is simplified, as only one dopant element is required. Hence,
the reduced fabrication costs could offset the lower efficiencies of Schottky-barrier thin-film solar cells.

Theoretical studies \cite{Mahala,Hamady2016}, which corroborate an earlier experimental study \cite{Xue}, suggest that \InGaN~Schottky-barrier solar cells with relatively high efficiency could be designed.  Anderson \textit{et al.} \cite{Anderson2017} investigated the efficiency of \InGaN~Schottky-barrier solar cells with periodic variation of the indium-to-gallium ratio. This involved {the solution of} both the frequency-domain Maxwell postulates in the optical regime and the carrier drift-diffusion equations using the commercial 
\cyan{finite-element  package} COMSOL (V5.2a), in order to simulate the efficiencies of a variety of designs. The efficiency  was found to increase significantly on the incorporation of periodic nonhomogeneity with a specific profile.

For the traditional amorphous-silicon $p$-$i$-$n$-junction solar cells,   the incorporation of a periodically nonhomogeneous intrinsic layer (i.e., $i$ layer), along with a metallic periodically corrugated back-reflector (PCBR), can improve overall efficiency by up to 17\% \cite{Anderson2016}. For a Schottky-barrier  solar cell made from \InGaN, the inclusion of these features was shown to increase the total efficiency by up to $26.8\%$ \cite{Anderson2017}. In neither case, however, was comprehensive optimization of the design parameters conducted. The improvements seen are  likely due to the following reasons:
\begin{itemize}
	\item[(i)] The excitation of guided wave modes, including surface-plasmon-polariton waves \cite{AndersonIEEE,AndersonSPIE,Polo_book} and waveguide modes \cite{Khaleque}, is made possible by the inclusion of the metallic PCBR \cite{SBS1983,HM1995,SolanoAO2013,Liu2015}. Both types of  phenomena intensify the optical electric field inside the photon-absorbing regions of the solar cell, which leads to an increase in the electron-hole-pair generation rate.
	\item[(ii)] The combination of a periodically nonhomogeneous semiconductor and a PCBR enables the excitation of an increased number of guided wave modes \cite{Faryadspie2013,Liu2015,Ahmad}. More pathways become available for the incident photons to be absorbed, 
	thereby increasing the charge-carrier-generation rate. 
	\item[(iii)] The drift-diffusion equations include terms pertaining to gradients in the electrical constitutive properties of the materials in the solar cell. The material nonhomogeneity will facilitate the separation of  electrons and holes, and it may also suppress recombination \cite{Kabir2012,Iftiquar_Book}.
\end{itemize}

The aim of this paper is to expand on the previous work on \InGaN~by providing a comprehensive optimization of the device parameters in order to maximize  efficiency. The optical calculations were undertaken using the the rigorous 
coupled-wave approach (RCWA) \cite{Polo_book}, while the electrical calculations were undertaken using COMSOL \cyan{(V5.3a)} \cite{comsol}.  Optical absorption could have been maximized if only optical models had been used,  but the missing influence of the varying electrical properties would have made optimization of efficiency impossible  \cite{Anderson2016}.  For example, if electrical modeling is omitted, the optical absorption can be maximized by minimizing the bandgap, but this would result in a solar cell with a small open-circuit voltage and therefore, quite likely,   low efficiency \cite{Civiletti}.

{The plan of this paper is as follows.} The design of the chosen solar cell is summarized in Sec.~\ref{Sec:SolarDesign}, with further details available elsewhere\cite{Anderson2016,THA_Photonics_West}. The optical and electrical constitutive properties used in the simulation are presented in Sec.~\ref{Sec:Materials}, while the computational models employed are described in Sec.~\ref{Sec:Compmodel}. Numerical results are presented in Sec.~\ref{Sec:Optimization}.  Closing remarks are presented in Sec.~\ref{Sec:closing}.

\section{Summary of the two-dimensional \cyan{(2-D)} model}
\subsection{Solar-cell design} \label{Sec:SolarDesign}
\begin{figure}[ht!]
\centering
\includegraphics[width=0.7\textwidth]{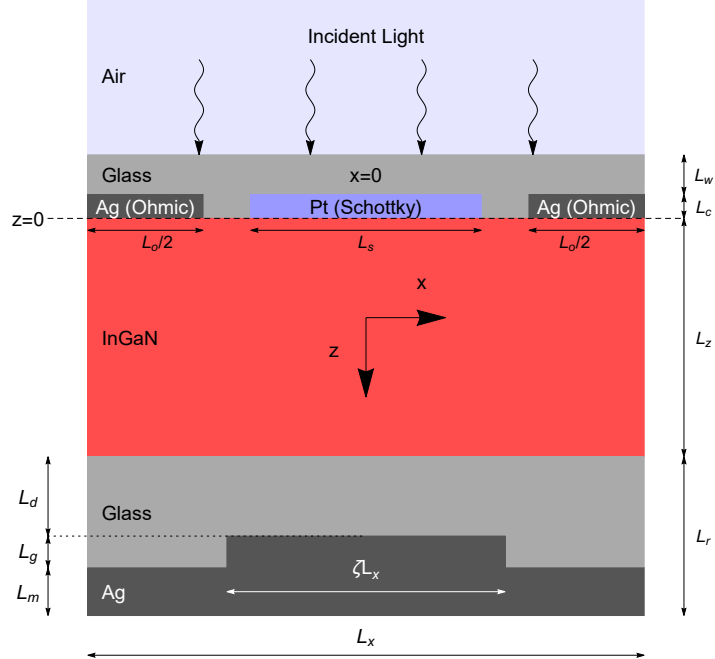}
\caption{Schematic illustration of the reference unit cell $\cal R$ of the   Schottky-barrier solar cell.}
\label{Fig:Schematic}
\end{figure}

 The model is  described in detail in   Ref.~\citenum{Anderson2017}.
  For the sake of completeness, a summary is included here. The   simulated Schottky-barrier solar cell is schematically illustrated in Fig.~\ref{Fig:Schematic}. As the solar cell is translationally invariant in the $y$~direction, the simulation is reduced to two dimensions  (i.e., the $xz$ plane) without approximation. In the remainder of this paper, the term \textit{width} refers to the extent along the  $x$ axis, whereas the term \textit{thickness} refers to the extent along the $z$ axis.
 
 The solar cell comprises a planar antireflection window, a layer containing  electrodes, a wafer of \InGaN, and a layer containing a \cyan{backreflector. }Each of these layers is of uniform thickness. Insolation occurs at normal incidence to the solar cell through the antireflection window, with the wave vector of the incident light aligned with the positive $z$ axis. 
 
The device is periodic along the $x$ axis with period $L_x$
and has a thickness $L_w+L_c+L_z+L_r$. The reference unit cell of the device is the region 
${\cal R} =\left\{(x,z)\vert -L_x/2 < x <L_x/2, -L_w-L_c\right.$  $\left.< z < L_z+L_r\right\}$. A planar antireflection window, made from 
\cyan{flint} glass \cite{Malitson65}, occupies the region $-L_w-L_c<z<-L_c$ in ${\cal R}$. The region $0<z<L_z$ is occupied by $n$-doped \InGaNn, forming both Schottky-barrier and ohmic junctions with the metal electrodes in the region {$-L_c<z<0$} in ${\cal R}$. For optical calculations, the ohmic contact and \cyan{backreflector were} assumed to be silver \cite{Palik}, while the Schottky-barrier contact was assumed to be platinum \cite{Werner2009}. It must be noted that the electrical properties of silver were not used. The Schottky-barrier electrode, of width $L_s$, is centered in $\cal R$ at $x=0$ along the $x$ axis. The two ohmic {electrodes}, each of width $L_o/2$, are centered at  $x=\pm \le L_x - L_o/2 \ri /2$  in $\cal R$. Note that $L_o + L_s < L_x$,  and so the electrodes are electrically isolated. It should also be noted that, due to the periodicity of the design, there are an equal number of ohmic and Schottky-barrier electrodes \cyan{in} the solar cell. The gaps between the {electrodes}, {$-L_c < z < 0$} and either $- (L_x - L_o)/2 < x < -L_s/2$ or $L_s/2 < x < (L_x - L_o)/2$, are occupied by flint glass.

The region $L_z <z<L_z +L_r$  in $\cal R$ contains both silver and flint glass. The  back-reflector is made of two silver slabs welded together. The first slab is optically thick and occupies the region $L_z+L_r-L_m<z<L_z+L_r$. The second slab
occupies the region $\lec{(x,z)\vert - \zeta L_x/2 < x <\zeta L_x/2,
L_z+L_d<z<L_z+L_r-L_m}\ric\subset \mathcal{R}$, where $\zeta \in(0,1)$ is the duty cycle. Thus,
$L_g=L_r-(L_d+L_m)$ is the corrugation height.
The remainder of the region $L_z <z<L_z +L_r$  in $\cal R$ is occupied by  flint glass which electrically insulates silver from \InGaN.
  
Absorption of the normally incident solar flux with AM1.5G spectrum \cite{AM1.5} was calculated by solving the frequency-domain Maxwell postulates \cite{Polo_book}. The semiconductor charge-carrier drift-diffusion equations model the \cyan{spatial distributions
of the
electron density and hole density} \cite{Jenny_Book, Fonash_book}. Because of the nonhomogeneity
of the semiconductor (i.e., \InGaNn), the effective dc electric field acting on 
{\begin{itemize}
\item[(a)] electrons includes a contribution from gradients in the electron affinity, and 
\item[(b)] holes includes contributions from gradients in both the electron affinity and the bandgap. 
\end{itemize}} Direct, mid-gap Shockley--Read--Hall, and Auger recombination were all included in our simulation. The current density $J$, which is averaged over either the Schottky-barrier {electrode} (or, identically, both of the ohmic {electrodes}), was calculated for a range of
\cyan{values of the} external biasing \cyan{voltage} $V_{ext}$.

\subsection{Material parameters} \label{Sec:Materials}
For a specific bandgap $\sfE_{g0}$, the fractional concentration of indium $\param$ is given by
\begin{align}
\param(z) = \frac{ b+(\sfE_g^{\tiny \mbox{GaN}}-\sfE_g^{\tiny \mbox{InN}})-\sqrt{4b\les{\sfE_{g0}(z)-\sfE_g^{\tiny \mbox{GaN}}}\ris+(b+\sfE_g^{\tiny \mbox{GaN}}-\sfE_g^{\tiny \mbox{InN}})^2}}{2 b},
\label{Eqn:EgComp}
\end{align}
where the bowing parameter $b = 1.43$~eV \cite{Anderson2017, Wu2003_2}, {and the bandgaps
$\sfE_g^{\tiny \mbox{InN}}=0.7$~eV and $\sfE_g^{\tiny \mbox{GaN}}=3.42$~eV.}

\subsubsection{Optical parameters}
The optical refractive index $\nopt$ of \InGaN~ depends on the free-spaced wavelength $\lambdao$ and 
was modeled using two equations. The real part of $\nopt$ is  {provided by the Adachi model as~\cite{Hamady2016}
\begin{align}
\mbox{Re}\lec \nopt(\param,\lambdao)  \ric  &= \mbox{Re} \lec \sqrt{ A_A(\param)\le\les\frac{\sfE_{g0}(\param)}{\sfE_{\gamma}(\lambdao)}\ris^2\les 2-\sqrt{ 1+\frac{\sfE_{\gamma}(\lambdao)}{\sfE_{g0}(\param)}}-\sqrt{1-\frac{\sfE_{\gamma}(\lambdao)}{\sfE_{g0}(\param)}}\ris+B_A(\param)\ri} \ric,
\end{align}
where $A_A(\param)$ and $B_A(\param)$ are
interpolated from the corresponding parameters for InN and GaN} provided in Table~\ref{Table:Parameters}.  The photon energy is denoted by $\sfE_{\gamma}(\lambdao)= 2 \pi \hbar \co /\lambdao$, where  $\hbar = 1.054571800  \times 10^{-34}$~m$^2$~kg~s$^{-1}$ is the reduced Planck   constant and $\co = 2.99 792 485\times10^{8}$~m~s$^{-1}$ is the speed of light in free space.

The imaginary part of the optical refractive index $\nopt$ was modeled as 
\begin{align}
\mbox{Im} \lec  \nopt(\param,\lambdao)  \ric =  \frac{\lambdao}{4\pi} \alpha_{\rm opt}(\param,\lambdao).
\end{align}
The absorption coefficient  $\alpha_{\rm opt}$ was
modeled by \cite{Hamady2016}
\begin{align}
\alpha_{\rm opt}(\param,\lambdao) = 10^5\sqrt{C(\param)\les{\sfE_{\gamma}(\param)-\sfE_{g0}(\param)}\ris+D(\param)\les{\sfE_{\gamma}(\param)-\sfE_{g0}(\param)}\ris^2} \,\, \mbox{nm}^{-1}\,,
\end{align}
wherein the  constants
\begin{align}
\left.\begin{array}{l}
C(\param)= \le 3.525-18.28\param+40.22\param^2-37.52\param^3+12.77\param^4 \ri \, \mbox{eV}^{-1} \\[5pt]
D(\param)  = \le -0.6651+3.616\param-2.460\param^2 \ri \, \, \mbox{eV}^{-2}
\end{array}\right\}
\end{align}
come from  interpolation of  parameters given by Brown \textit{et al.}~\cite{Brown2009}.

\subsubsection{Electrical Parameters}
The Schottky-barrier \cyan{work function} matched that of  platinum in our simulations. Thus, $\Phi = 5.93$~eV.\cite{Workfunction}
For a specific value of $\param$, the electrical properties were modeled using either quadratic or linear (i.e. Vegard's law \cite{Vegard}) interpolation of data for InN and GaN.

The electron affinity
\begin{align}
\chi_0(z) =\param(z) \chi^{\tiny\mbox{InN}} + \les 1-\param(z)\ris \chi^{\tiny\mbox{GaN}}-b\param(z)\les 1-\param(z)\ris\,,\quad
z\in(0,L_z)\,,
\label{Eqn:Chix}
\end{align}
was modeled using the same quadratic fit as the bandgap, where $\chi^{\tiny\mbox{InN}}$ and $\chi^{\tiny\mbox{GaN}}$ are the electron affinities of InN and GaN, respectively. 
 All other parameters   presented in the first column of Table \ref{Table:Parameters} were modeled using Vegard's law of linear interpolation.

\begin{table}[h!b]
\caption{Electronic data used for GaN and InN. The composition of \InGaN~was estimated using Eq.~(\ref{Eqn:EgComp}), with linear interpolation used to estimate data for the semiconductor-filled region $0<z<L_z$ with bandgaps not presented here in all cases, except for the electron affinity $\chi_0$ which uses Eq.~(\ref{Eqn:Chix}).
\label{Table:Parameters}}
\footnotesize
\rowcolors{1}{llightgray}{lightgray}
\makebox[\textwidth][c]{
\begin{tabular}{p{5cm}llll}\hline
	& Symbol & Unit & GaN & InN\\ \hline\hline
					Bandgap & $\sfE_{g}^*$ & $\mbox{eV}$			& $3.42$ & $0.7$\\ \hline\hline
Electron Affinity & $\chi_0$ & eV	& 4.1& 5.6\\
Density of States (Conduction Band) & $N_{C}$ & cm${}^{-3}$	& $2.3\times 10^{18}$ & $9.1\times 10^{17}$\\
Density of States (Valence Band) & $N_{V}$ &  cm${}^{-3}$	& $4.6\times 10^{19}$ &$ 5.3\times10^{19}$\\\hline
Electron Mobility 1 & $\mu_n^{(1)}$ &  {cm}$^2$~{V}$^{-1}$~{s}$^{-1}$ & $295$ &$1030$\\
Electron Mobility 2 & $\mu_n^{(2)}$ & {cm}$^2$~{V}$^{-1}$~{s}$^{-1}$ & $1460$ &$14150$\\
Caughey--Thomas Doping Power (Electrons) & $\delta_n$ &  & $0.71$ & $0.6959$\\
Caughey--Thomas Critical Doping Density (Electrons) & $N_n^{crit}$ & cm${}^{-3}$ & $7.7\times 10^{16}$ & $2.07\times 10^{16}$\\\hline
Hole Mobility 1 & $\mu_p^{(1)}$ & {cm}$^2$~{V}$^{-1}$~{s}$^{-1}$ & $3$ &$3$\\
Hole Mobility 2 & $\mu_p^{(2)}$ & {cm}$^2$~{V}$^{-1}$~{s}$^{-1}$ & $170$ &$340$\\
Caughey--Thomas Doping Power (Holes) & $\delta_p$ &  & $2$ & $2$\\
Caughey--Thomas Critical Doping Density (Holes) & $N_p^{crit}$ & cm${}^{-3}$ & $1\times 10^{18}$ & $8\times 10^{17}$\\\hline
Auger Recombination Factor (Electrons) &$C_n$ & cm${}^{6}$~s$^{-1}$ & $1.5\times 10^{-30}$ &  $1.5\times 10^{-30}$\\
Auger Recombination Factor (Holes) & $C_p$ & cm${}^{6}$~s$^{-1}$ & $1.5\times 10^{-30}$ &  $1.5\times 10^{-30}$\\
Direct Recombination Factor & $C_{rad}$& cm${}^{3}$~s$^{-1}$ & $1.1\times 10^{-8}$ & $2\times10^{-10}$\\\hline
Slotboom Reference Energy & $\sfE_{ref}$ & eV & $9\times 10^{-3}$ & $9\times 10^{-3}$\\
Slotboom Reference Concentration  & $N_{ref}$ & cm${}^{-3}$ & $1\times10^{17}$ &$1\times10^{17}$ \\
Conduction-Band Fraction & $\alpha_{oc}$ &   & $0.9$ & $0.9$ \\
Adachi Refractive-Index Parameter $A_A$ &$A_A$& & 9.31 & 13.55\\
Adachi Refractive-Index Parameter $B_A$ &$B_A$&  &3.03 & 2.05\\
\end{tabular}
}
\vspace{2mm}
\end{table}

The narrowing of the bandgap associated with doping was incorporated through the Slotboom model \cite{Slotboom1976}. An empirical low-field mobility model---called  either the Caughey--Thomas \cite{Hamady2016} or the Arora \cite{comsol} mobility model---was used for  the variations of the electron mobility and the hole mobility. Details of these models are available elsewhere \cite{Anderson2017,Hamady2016}.

The bandgap of \InGaN~ was taken to vary periodically in the thickness direction of the solar cell, described by
\begin{align}
\sfE_{g0}(z) = \sfE_{g}^*-A\le1-\lec \frac{1}{2}\les \sin\le   \frac{2\pi z}{L_p}-2\pi \phi\ri +1\ris\ric^{\alpha}\ri\,,\label{Eqn:Eg}
\end{align}
where $\sfE_{g}^*$ is the baseline (maximum) bandgap, $A$ is the  amplitude, $L_p=L_z/\tilde{\kappa}$ is the period with $\tilde{\kappa}>0$, $\phi$ is a phase shift, and $\alpha$ is a shaping parameter.

\subsection{Computational model} \label{Sec:Compmodel}
The problem of calculating the total efficiency of the solar cell was decoupled into two separate calculations. Firstly, the RCWA\cite{Polo_book}  was used to calculate the   spectrally integrated \cyan{photon-absorption} rate which is ideally equal
to the charge-carrier-generation rate. This  was then coupled with a 2D finite-element electronic model that was implemented in the \cyan{COMSOL  (V5.3a)} software package \cite{comsol}. In the remainder of this paper,  terms in small capitals are \cyan{{\sc COMSOL}} terms.

\subsubsection{Differential evolution algorithm}
The differential evolution algorithm (DEA) \cite{DEA} was used to optimize the solar-cell design. Given $\Nbar$ parameters in the optimization problem, an initial population $\#P_0$ of $N_P$ members in the parameter-search space  $\mathcal{S}\subset\mathbb{R}^{\Nbar}$ was chosen randomly, with a uniform distribution. After the cost function $C:\mathcal{S}\to\mathbb{R}$ of the problem had been evaluated at each of these points, the DEA {produced} a new population $\#P_1$ of $N_P$ members in the parameter search space $\mathcal{S}$ to test. This process 
was iterated until the \cyan{change in absorptance was less than 1\%} or until a set amount of time had passed.

By representing the current population $\#P_j$ as a matrix with each of its $N_P$ columns being vectors in $\mathcal{S}$, the next population can be written as
\begin{align}
\#P_{j+1} = \#{\sc M}_1 \circ \left[\#P_j + F (\#v^*\otimes\#1 - \#P_j) + 
F (\tilde{\#P}_j^{(1)} - \tilde{\#P}_j^{(2)})\right] + \#M_2 \circ \#P_j
\end{align}
where $\#v^*\in \mathcal{S}$ is the optimal parameter vector found {at that stage}; $\#1$ is the vector of \cyan{$1$'s}; $\otimes$ is the outer product; $\tilde{\#P}_j^{(1)}$ and $\tilde{\#P}_j^{(2)}$ are versions of $\#P_j$ where the columns have been randomly interchanged; the parameter $F \in (0,2]$ is the step size to be taken by the DEA at each iteration; $\#M_1$ and $\#M_2 = \#1\otimes\#1 - \#M_1$ are filter matrices of $1$'s and $0$'s generated by the DEA, with $\#M_1$ having approximately a fraction $C_R$ of \cyan{$1$'s}, where $C_R$ is termed the crossover fraction; and $\circ$ is the Shur product (elementwise multiplication).

The cost function $C$ was taken to be either the  efficiency $\eta$ or the optical short-circuit current density $J_{SC}^{Opt}$. The population number was set to $N_P = 30$, the crossover fraction was set to $C_R = 0.6$, and the step size 
$F$ was set to be randomly distributed in $\left[0.5,1\right]$ uniformly. Allowing $F$ to vary randomly with each iteration has been termed \textit{dither}, and has been shown to improve convergence for many {problems \cite{Swagatam2011}.}

\subsubsection{The RCWA algorithm}
{Suppose that the face $z=-L_c-L_w$ of the solar cell is illuminated by a normally incident plane wave with electric field phasor
\begin{equation}
{\underline E}_{\rm inc}(x,z,\lambdao)=
\frac{E_o}{\sqrt{2}} (\ux+\uy)\exp\left(i\frac{2\pi}{\lambdao}{z}\right)\,.
\end{equation}
As a result of
the metallic back-reflector being periodically corrugated,}
the $x$-dependences of the electric and magnetic field phasors {must be} represented by Fourier series everywhere as  
\begin{align}
{\underline E}(x,z,\lambdao) &= \sum_{n=-\infty}^{n=\infty} {\underline e}^{(n)}(z,\lambdao)\exp\left(i\kappa^{(n)}x\right),\quad \vert z\vert <\infty,\quad  \vert x\vert <\infty,
\intertext{and}
{\underline H}(x,z,\lambdao) &= \sum_{n=-\infty}^{n=\infty} {\underline h}^{(n)}(z,\lambdao)\exp\left(i\kappa^{(n)}x\right), \quad \vert z\vert <\infty,\quad  \vert x\vert <\infty,
\end{align}
where $i=\sqrt{-1}$, $\kappa^{(n)} =   n (\cyan{2\pi/L_x})$ and
 ${\underline e}^{(n)} = {e}_x^{(n)}\ux+{e}_y^{(n)}\uy+{e}_z^{(n)}\uz$ as well as ${\underline h}^{(n)}= {h}_x^{(n)}\ux+{h}_y^{(n)}\uy+{h}_z^{(n)}\uz$ are  Fourier coefficients. Likewise, the optical permittivity $\eps(x,z,\lambdao)
 =\epso\nopt^2(x,z,\lambdao)$ everywhere {has to be} represented by the Fourier series
 \begin{align}
\eps(x,z,\lambdao) &= \sum_{n=-\infty}^{n=\infty} {\eps}^{(n)}(z,\lambdao)\exp\left(i\kappa^{(n)}x\right),\quad \vert z\vert <\infty,\quad  \vert x\vert <\infty,
\end{align}
where $\epso$ is the permittivity of free space.

Computational tractability requires truncation {so that   $n \in \left\{-N_t,...,N_t\right\}$, $N_t\geq0$.}  Column vectors
\begin{align}
{\*e}_\sigma(z,\lambdao)=\les 
e_\sigma^{(-N_t)}(z,\lambdao),e_\sigma^{(-N_t+1)}(z,\lambdao),...,e_\sigma^{(N_t-1)}(z,\lambdao),e_\sigma^{(N_t)}(z,\lambdao)\ris^T,\quad\sigma\in\{x,y,z\},
\end{align}
and
\begin{align}
\*h_\sigma(z,\lambdao)=\les h_\sigma^{(-N_t)}(z,\lambdao),h_\sigma^{(-N_t+1)}(z,\lambdao), ...,h_\sigma^{(N_t-1)}(z,\lambdao),h_\sigma^{(N_t)}(z,\lambdao)\ris^T,\quad\sigma\in\{x,y,z\},
\end{align}
were set up, the superscript $T$ denoting the transpose.  Furthermore, the matrixes 
\begin{align}
\*K= 
{\rm diag}\les  \kappa^{(-N_t)},\, \kappa^{(-N_t+1)},\,..., 
\kappa^{(N_t-1)},\,\kappa^{(N_t)}\ris
\end{align}
and   
\begin{align}
\*{\boldsymbol \eps}(z,\lambdao)=
\les
\begin{array}{ccccc}
\eps^{(0)}(z,\lambdao) & \eps^{(-1)}(z,\lambdao) &  ... &   
\eps^{(-2N_t+1)}(z,\lambdao) & \eps^{(-2N_t)}(z,\lambdao)\\[5pt]
\eps^{(1)}(z,\lambdao) & \eps^{(0)}(z,\lambdao) &   ... &   
\eps^{(-2N_t+2)}(z,\lambdao) &\eps^{(-2N_t+1)}(z,\lambdao)\\[5pt]
 ... & ... & ...  &... & ... \\[5pt]
\eps^{(2N_t-1)}(z,\lambdao) & \eps^{(2N_t-2,\lambdao)}(z,\lambdao) &   ... &   
\eps^{(0)}(z,\lambdao) & \eps^{(-1)}(z,\lambdao)\\[5pt]
\eps^{(2N_t)}(z,\lambdao) & \eps^{(2N_t-1)}(z,\lambdao) &  ... &   
\eps^{(1)}(z,\lambdao) & \eps^{(0)}(z,\lambdao)
 \end{array}
\ris \label{Eq2.105}
\end{align}
were set up. The frequency-domain Maxwell curl postulates then {yielded} the
matrix ordinary differential equation
\begin{align}
\frac{d}{dz} \*f(z,\lambdao) =i\*P(z,\lambdao)\.\*f(z,\lambdao), \label{Eqn:diff}
\end{align}
where the $4(2N_t+1)$-column vector  
\begin{align}
\*f(z,\lambdao)=\les \begin{array}{c}
\*e_x(z,\lambdao)\\[3pt]
\*e_y(z,\lambdao)\\[3pt]
\*h_x(z,\lambdao)\\[3pt]
\*h_y(z,\lambdao)
\end{array}
\ris
\end{align}
and the $4(2N_t+1)\times4(2N_t+1)$ matrix  
\begin{eqnarray}
\nonumber
\*P(z,\lambdao)&=&\omega\les
\begin{array}{cccc}
\*0&\*0&\*0&\muo\*I\\[3pt]
\*0&\*0&- \muo\*I&\*0\\[3pt]
\*0& - \*{\boldsymbol \eps}(z,\lambdao) &\*0&\*0\\[3pt]
\*{\boldsymbol \eps}(z,\lambdao)&\*0&\*0&\*0
\end{array}\ris
\\[10pt]
&&\quad +\frac{1}{\omega}\les
\begin{array}{cccc}
\*0&\*0&\*0&
-  \*K\.\les\*{\boldsymbol \eps}(z,\lambdao)\ris^{-1}\.\*K\\[3pt]
\*0&\*0&\*0&\*0\\[3pt]
\*0& \muo^{-1}\*K\.\*K  &\*0&\*0\\[3pt]
\*0&\*0&\*0&\*0
\end{array}\ris 
\label{Eq2.109}
\end{eqnarray}  
contains $\muo$ as the permeability of free space, $\*0$ as the $(2N_t + 1)\times(2N_t+1)$ null matrix, and
$\*I$ as the $(2N_t + 1)\times(2N_t+1)$ identity matrix.

The solar cell was discretized along the $z$ axis  \cite{Polo_book}.  This effectively broke the domain $\cal R$ into a cascade of slices. Each  slice was homogeneous along the $z$ axis but it was either homogeneous or periodically nonhomogeneous along the $x$ axis. Equation (\ref{Eqn:diff}) 
was then solved  using a  stepping algorithm to give an approximation for $\*{f}$ in each slice. Finally, the Fourier coefficients of the $z$ components of the electric and magnetic field phasors were obtained from
\begin{align}
\left.\begin{array}{l}
 \*{e}_z(z,\lambdao) = -\omega^{-1}\les \*{\boldsymbol \eps}(z,\lambdao)\ris^{-1} \.
  \#{\breve{K}} \. \*{h}_y(z,\lambdao)\\[5pt]
\*{h}_z(z,\lambdao) =(\omega \muo)^{-1}  \#{\breve{K}} \.\*{e}_y(z,\lambdao)
\end{array}\right\}.
\end{align}
Thus, the electric field phasor was determined throughout the solar cell.  

{The spectrally integrated number of absorbed photons per unit volume per unit time is  given by
\begin{equation}
N_{ph}(x,z)= \frac{\etao}{\hbar }\int_{\lambdaomin}^{\lambdaomax}
\text{Im}\left\{\eps(x,z,\lambdao) \right\} \left\vert
\frac{{\underline E}(x,z,\lambdao)}{E_o}
\right\vert^2
S(\lambdao) \operatorname{d}\!\lambdao\,,\label{Eqn:Nabs}
\end{equation}
where $\etao=\sqrt{\muo/\epso}$ is the intrinsic impedance of free space
and $S(\lambdao)$ is the AM1.5G solar spectrum \cite{AM1.5}.
With the assumption that the absorption of every photon in the \InGaN~layer releases
an electron-hole pair, the
charge-carrier-generation rate can be calculated as 
\begin{align}
G(x,z) = N_{ph}(x,z)\label{Eqn:G}
\end{align}
everywhere in that layer.  Whereas ${\lambdaomin} = 350$~nm,  $\lambdaomax = 1240~\text{nm eV} /\sfE_{g,min}$ is the maximum wavelength that can contribute to the optical short-circuit current density 
\begin{align}
J_{SC}^{Opt} = q_e \frac{1}{L_x} \int_0^{L_z}\!\!\int_{-L_x/2}^{L_x/2} G(x,z) \operatorname{d}\!x\operatorname{d}\!z,\label{Eqn:JSCOpt}
\end{align}
 where $\sfE_{g,min}$ (in eV) is the minimum bandgap present in the solar cell and $q_e=1.6\times10^{-19}$~C is the elementary charge.} 

The integral in Eq.~(\ref{Eqn:Nabs}) was approximated using the trapezoidal rule \cite{Jaluria} with sampling at wavelengths spaced at \cyan{$2$-nm} intervals. The integral in \cyan{Eq.~(\ref{Eqn:JSCOpt})} was also approximated using the trapezoidal rule. The sampling resolution was regular in both directions, with $\delta x = L_x/100$, and $\delta z = L_z/200$.

The optical short-circuit current density   provides a rough benchmark for the device efficiency, and is  used by many optics researchers who simulate solar cells \cite{Anttu2017}. However, as recombination is neglected, $J_{SC}^{Opt}$  is necessarily larger than the actually attainable short-circuit current density $J_{SC}$, {which is} the electronically simulated current density that flows when the solar cell is illuminated and no external bias is applied (i.e., when $V_{ext}= 0$). For the results presented here,  calculating only $J_{SC}^{Opt}$ would have been inadequate as the electrical constitutive properties were also  significantly varied.

\subsubsection{Adaptive-$N_t$ implementation}
The calculated value of \JscOpt~varies with $N_t$. An adaptive method was implemented to estimate when $N_t$ is sufficiently large. \cyan{Equation~(\ref{Eqn:Nabs})} was evaluated using the trapezoidal rule \cite{Jaluria}. At the first value of $\lambdao$ sampled, $\lambdao = \lambdaomin$, \JscOpt~ was calculated with $N_{t_1}$ and $N_{t_2} = N_{t_1}+\cyan{2}$. If the magnitude {$\Delta{\JscOpt}=\vert\JscOpt(N_{t_2})-\JscOpt(N_{t_1})\vert$
 of the difference between the two calculated values of \JscOpt~ was greater than a specified tolerance, then $N_{t_1}$ was set equal to $N_{t_2}$ and $N_{t_2}$ was increased by two. This iterative procedure was continued until $\Delta{\JscOpt}$ was less than the specified tolerance for} two subsequent comparisons. A maximum value of $N_t \leq 100$ was enforced in order to force to the calculation to terminate within a reasonable duration. After a successful calculation, the \cyan{next  value} of $\lambdao$ was selected, and $N_{t_1} = 2 \ceil{N_{t_2}/4}$ and $N_{t_2} = N_{t_1} +\cyan{2}$ were chosen for the next calculation, where $\ceil{}$ is the ceiling function.

\subsubsection{Solution of  drift-diffusion equations}
The charge-carrier-generation rate $G(x,z)$, calculated using the RCWA, was processed using an external Matlab\texttrademark\, code and then used as the input, \cyan{via  {\sc User-Defined Generation},} for the COMSOL electrical model. Recombination was incorporated via {\sc Auger}, {\sc Direct} and {\sc Trap-Assisted (Midgap Shockley--Read--Hall)} phenomena, using parameters as provided in Table~\ref{Table:Parameters}. 

Due to the symmetry in the simulation, only the right half of the domain $\cal R$ (i.e., $0\leq x\leq L_x/2$) was electrically simulated, with {\sc Insulator Interfaces} applied down that line of \cyan{symmetry. }

{\sc Fermi-Dirac carrier statistics} were employed along with a {\sc Finite volume (constant shape function)} discretisation, as this inherently conserves current throughout the solar cell~\cite{comsol}. COMSOL utilizes a Scharfetter-Gummel upwinding scheme. The {\sc Free triangular}, {\sc Delaunay} mesh has a maximum element size of $15$~nm. Further details can be found in Ref.~\citenum{Anderson2017}.

The {\sc Semiconductor} module of COMSOL was used to calculate the current densities  flowing through the {\sc ohmic}, and therefore also {\sc Ideal Schottky}, {electrodes}. A prescribed external voltage $V_{ext}$ was applied between these {electrodes}. The current density flowing through the Schottky-barrier electrode was modeled using {\sc Thermionic currents}, with standard Richardson coefficients of $A^*_n = 110$~A~K$^{-2}$~cm$^{-2}$, and $A^*_p = 90$~A~K$^{-2}$~cm$^{-2}$\cite{Hamady2016,comsol}. By {sweeping $V_{ext}$} from 0~V up to a value where {$J_{SC}$} drops to zero, the $J_{SC}$-$V_{ext}$ curve was produced. This enabled calculation of the maximum
\cyan{attainable value of the efficiency $\eta$.}

\section{Numerical simulation results}\label{Sec:Optimization}
\subsection{Optimization Study}
\begin{table}
	\caption{Summary of parameters used for simulation. \label{Table:Param}}
	\centering
	\rowcolors{1}{llightgray}{lightgray}
	\begin{tabular}{cp{7cm}ll}
		Parameter 	& Name & Symbol & Value/Range\\\hline\hline
		$\checkmark$&Device Period & $L_x$ & $[100, 1500]$~nm\\
		$\checkmark$&Thickness of $n$-doped \InGaN~Layer & $L_z$ & $\{20,800\}$~nm\\
		& Minimum Thickness of Insulation Window & $L_d$ & $40$~nm\\
		$\checkmark$&Corrugation Height & $L_g$ & $[1,200]$~nm\\
		$\checkmark$&Corrugation Duty Cycle & $\zeta$ & [0.01,0.99]\\
		&Minimum Thickness of Metal & $L_m$ & $100$~nm\\
		&{Electrode}-Region Thickness & $L_c$ & $50$~nm\\
		$\checkmark$&Ohmic-{Electrode} Width & $L_o$ & $50+$~nm\\
		$\checkmark$&Schottky-barrier {Electrode} Width & $L_s$ & $50+$~nm \\
		& Antireflection-Window Thickness & $L_w$ & $75$~nm\\
		\hline
		$\checkmark$&Pre-doping Bandgap & $\sfE_g^*$ & $[0.7,3]$~eV\\
		$\checkmark$&Bandgap Nonhomogeneity Amplitude & $A$ & $[0, 3.42 - E^*_g]$~eV\\
		$\checkmark$&Bandgap Nonhomogeneity Phase & $\phi$ & $[0,1]$\\
		$\checkmark$&Bandgap Nonhomogeneity Shaping Parameter & $\alpha$&  $[2^{-4}, 2^{8}]$\\
		$\checkmark$&Bandgap Nonhomogeneity Ratio &$\tilde{\kappa}$
		 & $[0.01,5.5]$
		\\
	\end{tabular}
\end{table}

The defined problem has 15 parameters, shown in Tables~\ref{Table:Param}, all of which  influence the 
charge-carrier-generation rate $G(x,z)$ and the efficiency $\eta$ of the solar cell. The choice of four of these parameters was guided by either physical constraints, or they were found to strongly affect the optical response of the solar cell 
\cyan{while  affecting} the electrical characteristics only indirectly, i.e., by changing the spatial profile of the \cyan{charge-carrier-generation} rate. 
\begin{itemize}
\item A smaller value of $L_d$ was seen to further increase the number of photons absorbed in the \InGaNn; however, a minimum value of $L_d = 40$~nm was {chosen: (i) to model a solar cell where the electrical insulation of the backreflector is maintained, and (ii)} to provide a suitable surface for \InGaN~deposition.	
\item  The effect of the minimum thickness $L_m$ of the metallic layer has been omitted from the results. This is because, given a sufficiently thick silver layer, only a trivial amount of the incident light will be transmitted through the solar cell. {With $L_m$  chosen to be} more than   twice the penetration depth of silver across the majority of the AM1.5G {spectrum, changes} in solar-cell performance \cyan{for  modest} perturbations in $L_m$ from this value are minimal.
\item A smaller value of $L_c$ was seen to further increase the number of photons absorbed in the \InGaNn. A minimum value of $L_c = 50$~nm was chosen so the contacts are more physically realistic: thinner contacts would 
\cyan{likely suffer from  uneven} deposition and \cyan{exhibit} increased series resistance.
\item A strong peak  in efficiency was seen when $L_w \approx 75$~nm, irrespective of the other parameters. This corresponds to the glass coating acting as a quarter-wavelength antireflection coating. 
\end{itemize}
For all {data reported here,} the values $L_d= 50$~nm, $L_m= 100$~nm, $L_c= 50$~nm, and {$L_w = 75$~nm} were fixed.

The remaining 11 parameters were allowed to vary within the following ranges:  $L_x \in [100,1500]$~nm, $L_z \in [20,800]$~nm, $L_g \in [1,200]$~nm,  $\zeta \in [0.01,0.99]$,  $L_o \in [50,L_x - L_s]$~nm, $L_s \in [50,L_x - L_o]$~nm, $\sfE_{g}^* \in [0.7,3]$~eV, $A\in[0, 3.42 - \sfE_g^*]$~eV, $\phi\in[0,1)$, $\alpha\in[2^{-4},2^8]$, and $\tilde{\kappa}\in[0.01,5.5]$. These ranges, along with the chosen values of the optical parameters, are summarized in Table~\ref{Table:Param}. 

The maximum obtained efficiency   \cyan{$\eta=11.13\%$} was found at: $L_x = 694$~nm, $L_g = 125$~nm, $\zeta = 0.012$, $L_o/L_x = 0.16$, $L_s/L_x = 0.01$, $A = 0.74025$, $\cyan{\tilde{\kappa}} = 3.05$, $\phi = 0.711$, $\alpha = 13.3614$, \cyan{$\sfE_g^* = 1.17$~eV}, and \cyan{$L_z = 735$~nm}. These values were computed after 10 DEA population evolutions, and provide an estimate for the maximum efficiency, as well as allowing the upcoming interpretation of the results. Further iterations would provide greater confidence in the conclusions, at the cost of greater computation time.
 
\subsubsection{Results of Optimization Study}
\begin{figure}[ht!]
	\centering
	\begin{tabular}{cc}
\includegraphics[width=0.95\textwidth]{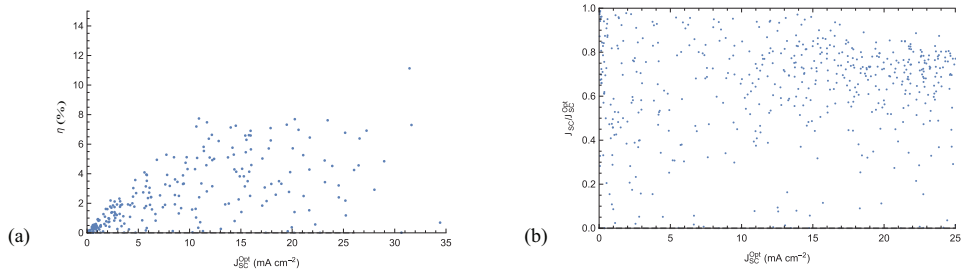} 
\end{tabular}
\vspace{5mm}
\caption{(a) Comparison of the  solar-cell efficiency {$\eta$ to the optical short-circuit current density $J_{SC}^{Opt}$, showing that an increase in the latter does not necessarily translate into an increase in the former. (b) Relationship between optical short--circuit current density and simulated short--circuit current density. For all data points, $L_d= 50$~nm, $L_m= 100$~nm, $L_c= 50$~nm, and $L_w = 75$~nm were fixed, but the other parameters were varied.}
\label{Fig:etavsJscOpt}}
\end{figure}

Figure \ref{Fig:etavsJscOpt} shows $\eta$ plotted against \JscOpt, with each data point corresponding to one DEA population member. The maximum value of \JscOpt~was calculated to be $37.042$~\mAcm2, but the maximum efficiency was observed to occur at $\JscOpt=31.442$~\mAcm2. Whereas a larger value of  \JscOpt~increases the likelihood of obtaining a high value of $\eta$, the former  does not predict the latter. Indeed, the designs with values of $\JscOpt > 25$~\mAcm2 produced efficiencies ranging from less than $1\%$ up to over $11\%$ in \cyan{Fig.~\ref{Fig:etavsJscOpt}(a)}. In part, this is caused by a device with a large optical short-circuit current density $J_{SC}^{Opt}$ not automatically producing a large short-circuit current density $J_{SC}$. This behavior is shown by the 
\cyan{droop} at higher values of $J_{SC}^{Opt}$ in 
\cyan{Fig.~\ref{Fig:etavsJscOpt}(b)}: as the optical short-circuit current density increases,  recombination in the  solar cell increases. These observations highlight the importance of conducting full {optoelectronic} simulations when modeling solar cells, especially when parameters with a strong electrical effect, such as the bandgap, are allowed to vary.

\subsubsection{Details of Optimization Study}
\begin{figure}
	\begin{tabular}{cc}
		 \includegraphics[width=0.9\textwidth]{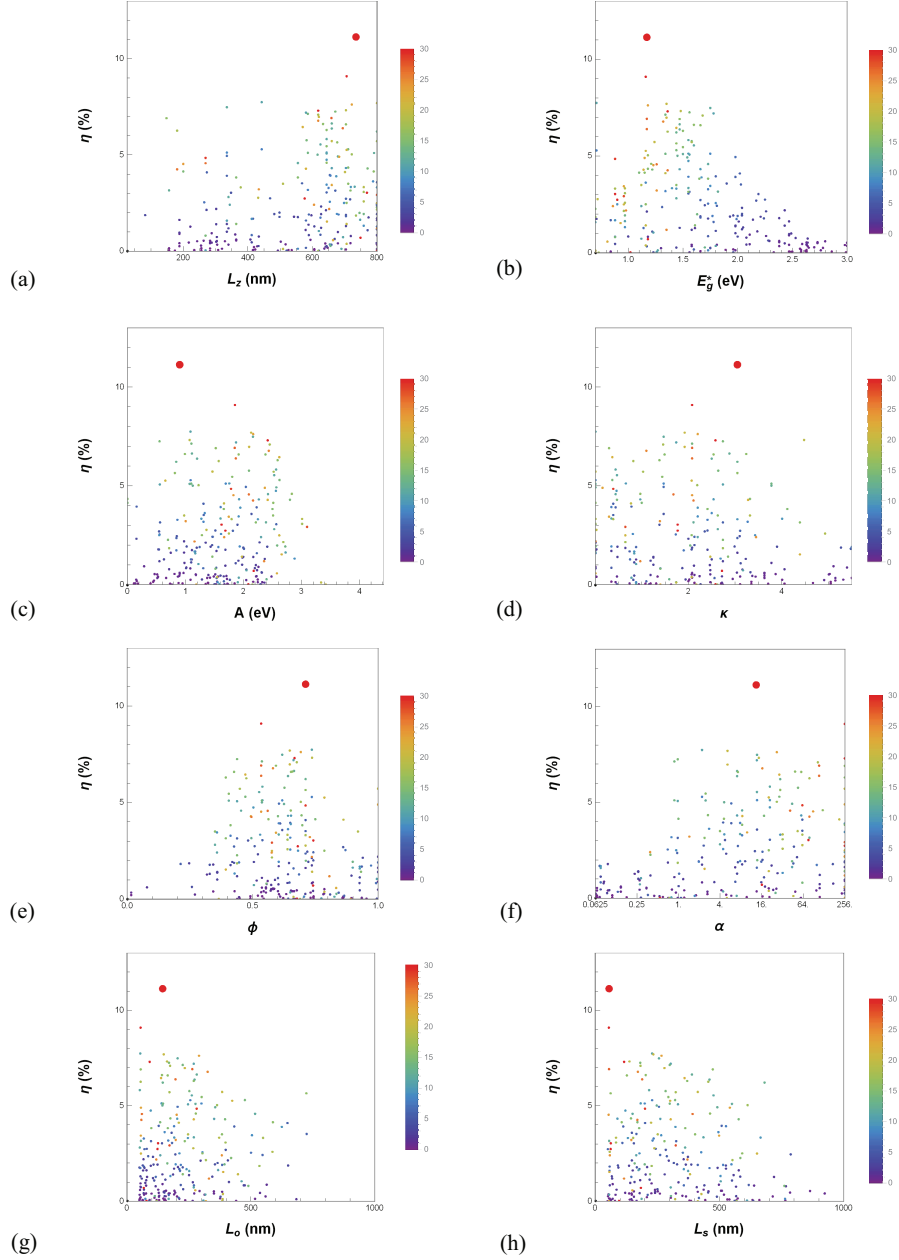}
	\end{tabular}
	\caption{Results projected onto the plane containing $\eta$~and (a) $L_z$, (b) $\sfE_g^*$, (c) $A$, (d) $\tilde{\kappa}$,
		(e) $\phi$, (f) $\alpha$, (g) $L_o/L_x$, and (h) $L_s/L_x$ for  a solar cell
		with $L_d= 50$~nm, $L_m= 100$~nm, $L_c= 50$~nm, and $L_w = 75$~nm. The large  points highlight the location of the 
		\cyan{device with the maximum efficiency}.} \label{Fig:Results}
\end{figure}

\begin{figure}[t!]
		\begin{tabular}{cc}
		\includegraphics[width=0.9\textwidth]{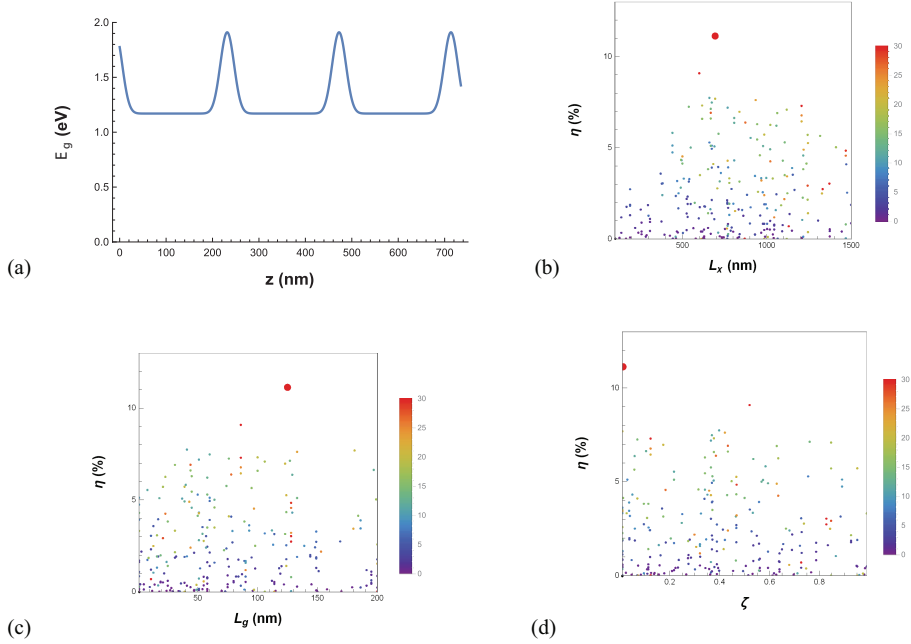} 		
	\end{tabular}

	\caption{(a) Bandgap of most efficient design. Rest, as Fig.~\ref{Fig:Results} except that the results are projected onto the plane containing $\eta$~and (b) $L_x$, (c) $L_g$, and (d) $\zeta$.} \label{Fig:Results2}
	
\end{figure}

The results from the  optimization study show how the different parameters affect {the  efficiency $\eta$} of the solar cell.  
\cyan{Figures}~\ref{Fig:Results} and \ref{Fig:Results2} show the projections of the entire parameter space onto the sets of axes containing the efficiency  and each of the optimization parameters. Parameters which have a strong effect on the  efficiency have most of their points strongly clustered around the design with the highest efficiency. Each point is colored with the value of $J_{sc}^{Opt}$.

In \cyan{Fig.~\ref{Fig:Results}(a)}, the thickness of the solar cell is seen to have a {moderate effect} on the resulting efficiency. The peak visible around $L_z \approx 700$~nm sees light clustering, with some reasonably efficient solar cells, with $\eta >7\%$, also produced when $L_z$ is less than half the optimal value. \cyan{Figure~\ref{Fig:Results}(b)} shows that \cyan{$\sfE_g^*$} strongly affects the resulting solar cell efficiency. The peak around $1.2$~eV lie in the region predicted by the Shockley-Queisser limit. Solar cells with narrower {bandgaps do produce solar cells with higher optical short-circuit current densities, but the reduction in open-circuit voltage  dramatically reduces} the efficiency. 
\cyan{Figure~\ref{Fig:Results}(c)} shows that a nonzero  amplitude $A$ can substantially increase the  efficiency. While the structure of the peak is not well resolved, all solar cells with $A<0.5$~eV had efficiencies less than half that of the maximum attained efficiency. Further increase of $A$ beyond $1$~eV slowly reduces the attainable efficiency. \cyan{Figures~\ref{Fig:Results}(g) and \ref{Fig:Results}(h)} indicate that a small Schottky-barrier \cyan{electrode and} a slightly larger ohmic {electrode} are required to maximize the efficiency.

A previous study \cite{Anderson2017} has suggested that optimal values of $\tilde{\kappa}$ are integer multiples of $1.5$, which conclusion is not contradicted by the data; see \cyan{Fig.~\ref{Fig:Results}(d).} The strong peak around $\phi = 0.75$  in 
\cyan{Fig.~\ref{Fig:Results}(e)} is also in line with previous work \cite{Anderson2016, Anderson2017}. At this value of $\phi$, a wide bandgap is produced in near to the electrodes. This seems of paramount importance for producing high efficiency solar cells with nonhomogeneous bandgaps. The best bandgap profile is shown in \cyan{Fig.~\ref{Fig:Results2}(a).}

Finally, in Fig.~\ref{Fig:Results2}, the effects of the \cyan{PCBR} are shown. A period of $L_x > 500$~nm is seen to significantly increase solar cell efficiency. This because the majority of short-wavelength light is absorbed far from the 
\cyan{PCBR} and so scattering effects are minimal. The grating amplitude and duty cycle 
\cyan{are} not seen to \cyan{have   strong effects} on the 
\cyan{solar-cell efficiency---the} points do not cluster strongly in \cyan{Figs.~\ref{Fig:Results2}(c) and \ref{Fig:Results2}(d).}

\subsection{Detailed study }

\begin{figure}[h!t]
	\begin{tabular}{cc}
		\includegraphics[width=0.9\textwidth]{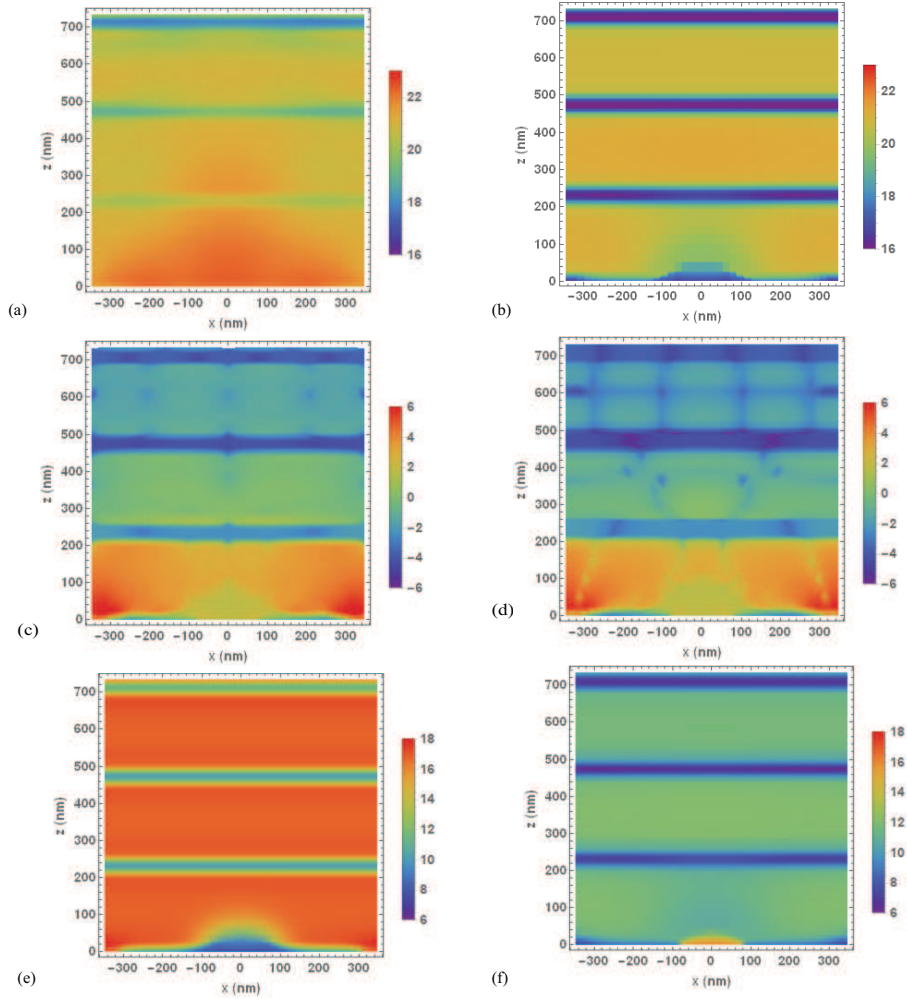} 		
	\end{tabular}
	\caption{Properties of the optimal solar cell at short--circuit condition: (a) generation rate in log$_{10}($cm$^{-3}$s$^{-1}$), (b) recombination rate in log$_{10}($cm$^{-3}$s$^{-1}$), (c) current density in log$_{10}($mA cm$^{-3}$), (d) current density in z--direction in log$_{10}($mA cm$^{-3}$), (e) electron density in log$_{10}($cm$^{-3}$), and (f) hole density in log$_{10}($cm$^{-3}$).}\label{Fig:Details}
\end{figure}

\begin{figure}
	\centering
		\includegraphics[width=0.6\textwidth]{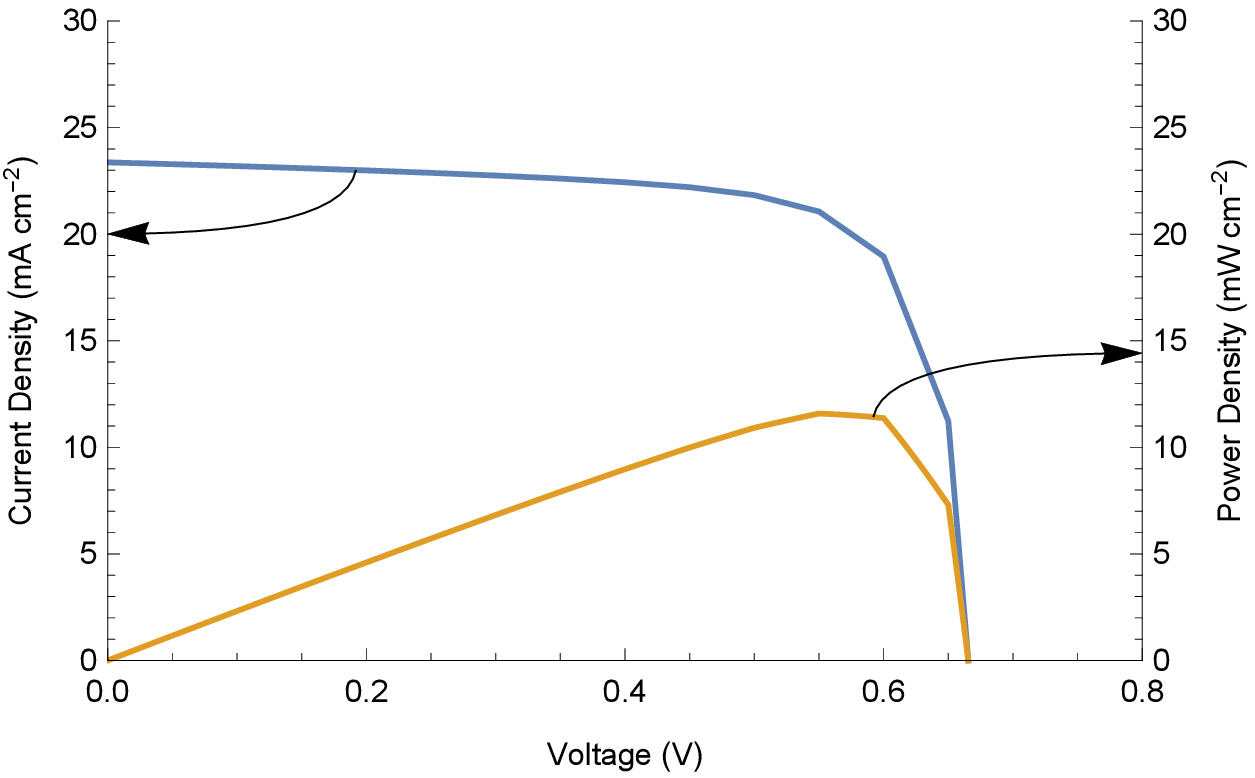}
		\caption{The resulting JV curve for the most efficient solar cell design.} \label{Fig:JV}
	\end{figure}
	
The simulated values of the major variables for the highest-efficiency device are shown in Fig.~\ref{Fig:Details}. The first two subfigures, (a) and (b), show the  generation rate and the recombination rate within the solar cell when the external voltage is zero. Light is incident from below
\cyan{in Fig.~\ref{Fig:Details}}. The bands of low generation and low recombination at $z\approx\{230, 570, 710\}$~nm correspond to the locations where the bandgap perturation is large, i.e. the bandgap is much wider here. The majority of the photons absorbed in the first $~200$~nm of the solar cell are collected before recombining, \cyan{whereas} the majority of those absorbed in the rear $~500$~nm recombine. The exception to this \cyan{trend} are the photons which are absorbed where the bandgap peaks. These are quickly swept \cyan{out} of these regions by the effective electric field produced by the gradient in the bandgap and electron affinity.

\cyan{Figures~\ref{Fig:Details}(c) and \ref{Fig:Details}(d)} show the current density produced in the solar cell. By comparing the two figures, it is seen that in the region of the narrow ohmic contact, at the outer edges of the plots, the current density is strongly perpendicular to the contact. The current density in the vicinity of the Schottky barrier, at the center of the plots, is lower, but \cyan{it is}
also perpendicular to the contact. The current density towards the back of the solar cell is dramatically lower, supporting the earlier analysis that the majority of the excited carriers in this region recombine.

\cyan{Figures~\ref{Fig:Details}(e) and \ref{Fig:Details}(f)} show the electron and hole densities at the short--circuit condition. The electrons are the majority carrier \cyan{in the majority} of the solar cell. In the vicinity of the Schottky-barrier junction, there is a high concentration of 
\cyan{holes.} In the regions where the bandgap is large, both carrier densities are very low, which has the effect of drastically reducing recombination in these areas. Unfortunately, it also acts to limit current flow across these bands: it may therefore be beneficial in future studies to 
include \cyan{lower-bandgap pathways} to increase \cyan{charge-carrier} extraction from further back in the solar cell.

Finally, Fig.~\ref{Fig:JV} shows the resulting JV curve for the optimal device. The short--circuit current is 21.79~mA~cm$^{-2}$, the open--circuit voltage is 0.683~V, and the fill factor is 0.7479.

\section{Closing remarks}\label{Sec:closing}
A combined optoelectronic model was developed to enable to optimization of \InGaN~based Schottky-barrier solar cells. These solar cells possessed a periodically corrugated back-reflector (PCBR), a layer of \InGaN, metallic {electrode}s, and an \cyan{antireflection} coating. The optimization was conducted using the Differential Evolution Algorithm. The AM1.5G solar spectrum was used to illuminate the solar cell at normal incidence.

With a solely optical model, it was shown that the optical short--circuit current density \JscOpt of the design is strongly dependent on the thicknesses of the materials that are applied to the surface. A $75$~nm thick layer of \cyan{flint} glass acts as a quarter wavelength \cyan{antireflection coating \cite{Chen2001,Moghal2012}}, maximizing \JscOpt. Minimizing the thickness of the front metallic electrodes also maximizes \JscOpt by reducing reflection.

An optimization study, using the differential evolution algorithm and a full optoelectronic model, produced a design for an \InGaN~based Schottky-barrier solar cell with a simulated efficiency of $11.13\%$. This design included a periodically nonhomogeneous bandgap, with just over 
\cyan{three} full periods. The minimum bandgap was $1.17$~eV \cyan{and the maximum bandgap was $1.91$~eV.} The phase of the periodic nonhomogeneity was such that the \InGaN~close to the {electrode}s has a wide bandgap.

While experimental work is needed to test the veracity of the models employed, it had been shown that \InGaN~based Schottky-barrier solar cells with a high efficiency may be producible if a periodic material nonhomogeneity is included.

\acknowledgments

This paper is  based in part on a paper entitled, ``Optimal indium-gallium-nitride Schottky-barrier thin-film solar cells," presented at the SPIE
conference Next Generation Technologies for Solar
Energy Conversion VIII, held August 5 to 11, 2017 in
San Diego, California, United States \cite{Proceedings}. The authors thank F. Ahmad (Pennsylvania State University) for assistance with Figs. 2--5.
The research of T.H. Anderson and P.B. Monk is partially supported by the US National Science Foundation under grant number DMS-1619904. The research of  A. Lakhtakia is partially supported by the  US National Science Foundation under grant number DMS-1619901.  A. Lakhtakia also thanks the Charles Godfrey Binder Endowment at the Pennsylvania State University for ongoing support of his research.

\noindent  \textbf{Tom H. Anderson} received his M.Sc. from the University of York in 2012, for work on modeling of non-local transport in Tokamak plasmas. He received his Ph.D. from the University of Edinburgh, UK, in 2016 for his thesis entitled \textit{Optoelectronic Simulations of Nonhomogeneous Solar Cells}. He is presently affiliated with the University of Delaware. His current research interests include optical and electrical modeling of solar cells, numerics, plasma physics, and plasmonics.\\

\noindent  \textbf{Akhlesh Lakhtakia} is Evan Pugh University professor and the Charles Godfrey Binder professor of engineering science and mechanics at the Pennsylvania State University. His current research interests include surface multiplasmonics, solar cells, sculptured thin films, mimumes, bioreplication, and forensic science. He has been elected a fellow of OSA, SPIE, IoP, AAAS, APS, IEEE, RSC, and RSA. He received
the 2010 SPIE Technical Achievement Award and the 2016 Walston Chubb Award for Innovation.\\

 \noindent   \textbf{Peter B. Monk} is a Unidel professor in the Department of Mathematical Sciences at the University of Delaware. He is the author of 
\textit{Finite Element Methods for Maxwell's Equations} and co-author with F. Cakoni and D. Colton of 
\textit{The Linear Sampling Method in Inverse Electromagnetic Scattering}.\\

\end{document}